\documentclass[
amsmath,
amssymb,
aps,
preprint,
superscriptaddress,
]{revtex4-2}
\usepackage{graphicx}
\usepackage{bm}
\usepackage{color}
\usepackage{soul}
\usepackage{lipsum}

\draft 

\begin{document}
\title{Machine-learning-enhanced quantum sensors for accurate magnetic field imaging} 
\author{Moeta Tsukamoto}
\email{moeta.tsukamoto@phys.s.u-tokyo.ac.jp}
\affiliation{Department of Physics, The University of Tokyo, Bunkyo-ku, Tokyo, 113-0033, Japan}
\author{Shuji Ito}
\affiliation{Department of Physics, The University of Tokyo, Bunkyo-ku, Tokyo, 113-0033, Japan}
\author{Kensuke Ogawa}
\affiliation{Department of Physics, The University of Tokyo, Bunkyo-ku, Tokyo, 113-0033, Japan}
\author{Yuto Ashida}
\affiliation{Department of Physics, The University of Tokyo, Bunkyo-ku, Tokyo, 113-0033, Japan}
\affiliation{Institute for Physics of Intelligence, The University of Tokyo, Bunkyo-ku, Tokyo, 113-0033, Japan}
\author{Kento Sasaki}
\email{kento.sasaki@phys.s.u-tokyo.ac.jp}
\affiliation{Department of Physics, The University of Tokyo, Bunkyo-ku, Tokyo, 113-0033, Japan}
\author{Kensuke Kobayashi}
\email{kensuke@phys.s.u-tokyo.ac.jp}
\affiliation{Department of Physics, The University of Tokyo, Bunkyo-ku, Tokyo, 113-0033, Japan}
\affiliation{Institute for Physics of Intelligence, The University of Tokyo, Bunkyo-ku, Tokyo, 113-0033, Japan}
\affiliation{Trans-scale Quantum Science Institute, The University of Tokyo, Bunkyo-ku, Tokyo, 113-0033, Japan}
\maketitle

\date{\today}

\maketitle

Local detection of magnetic fields is crucial for characterizing nano- and micro-materials and has been implemented using various scanning techniques~\cite{Marchiori2021,Moler2017} or even diamond quantum sensors~\cite{Taylor2008, Balasubramanian2008, Rondin2014,Degen2017,Casola2018}.
Diamond nanoparticles (nanodiamonds) offer an attractive opportunity to achieve high spatial resolution because they can easily be close to the target within a few 10 nm simply by  attaching them to its surface~\cite{Foy2020}.
A physical model for such a randomly oriented nanodiamond ensemble (NDE) is available~\cite{Foy2020}, but the complexity of actual experimental conditions still limits the accuracy of deducing magnetic fields.
Here, we demonstrate magnetic field imaging with high accuracy of 1.8~$\mu$T combining NDE and machine learning without any physical models. 
We also discover the field direction dependence of the NDE signal, suggesting the potential application for vector magnetometry and improvement of the existing model. 
Our method further enriches the performance of NDE to achieve the accuracy to visualize mesoscopic current and magnetism in atomic-layer materials ~\cite{Novoselov2005,Zhang2005,Cao2018,Burch2018,Mak2019} and to expand the applicability in arbitrarily shaped materials~\cite{Casola2018}, including living organisms~\cite{Schirhagl2014,Sage2013}. 
This achievement will bridge machine learning and quantum sensing for accurate measurements.

The nitrogen-vacancy (NV) center in diamond [Fig.~1(a)] is a point defect where a nitrogen atom replaces a carbon atom in the lattice accompanied by a neighboring vacancy.
By measuring its photoluminescence intensity while irradiating the laser and microwaves, NV's electron spin resonance can be detected, which is called optically detected microwave resonance (ODMR)~\cite{Barry2020}.
As the NV's spin level splits against the magnetic field in the direction of the NV symmetry axis $(111)$ due to the Zeeman effect~\cite{Loubser1978}, the determination of the ODMR frequency serves as quantum sensing of the field~\cite{Taylor2008}.
To obtain a nanoscale spatial resolution, we must attach the NV centers close to the sample within a few 10 nm~\cite{scholten2021}.
For this purpose, scanning techniques using diamond probes~\cite{Balasubramanian2008,Gross2017} or attachments of micro-fabricated diamond pieces~\cite{Du2017,Wang2022} are used.

The nanodiamonds (ND) could be an alternative to realize a similar adjacency~\cite{Foy2020}.
ND is a diamond crystal with a diameter of 5--200~nm harboring NV centers~\cite{Schirhagl2014}.
Since it enables us to adhere NV centers to any materials with arbitrary shapes by simply dropping NDs dispersed in a liquid, it has been applied to the measurement of electronic devices~\cite{Foy2020}.

Figure~1(b) shows our setup~\cite{Tsukamoto2021} to precisely measure the magnetic field dependence of the ODMR spectrum of NDE. A three-axis Helmholtz coil is used to adjust the magnetic field with a precision of $\pm$0.12~$\mu$T.
After the ODMR measurement, the magnetic field exactly at the sample location is determined with a tesla meter (Lake Shore Cryotronics F71).
Hereafter, this field value is referred to as the true magnetic field.

The magnetic field dependence of the ODMR spectrum of NDE (Method) on a cover glass is shown in the image plot of Fig.~1(c). This result is obtained from a total of one million NV centers contained in about 30,000 NDs.
The spectrum is measured at 751 points from 6~$\mu$T to 2286~$\mu$T with a field applied in the optical (z-axis) direction.
The spectrum shape becomes broadened with two dips as the field increases due to Zeeman splitting, and the ODMR contrast degrades.
This behavior reflects the fact that the crystal axes of each ND are random, resulting in different resonance frequencies of NV centers.

The spectrum is deformed for many reasons, such as the excitation efficiency, the light collection efficiency, the direction of the microwave and light polarization, etc., making it difficult to build an effective model for accurate magnetic field determination.
For example, we fit the spectrum obtained at the true magnetic field of 547.1$\pm$0.12~$\mu$T with an existing physical model that considers the experimental configuration (Method)~\cite{Foy2020}, which yields an estimated field of 582.2$\pm$4.4~$\mu$T, a statistically inaccurate value. 
This observation exemplifies the problem of hindering accurate local magnetometry using NDE. 
While NV centers are known to show very high sensitivity to the magnetic field ---how sensitive the signal changes depending on the field modulation---, the relevance of the accuracy achieved by them, that is, trueness of the measured value, has been less emphasized. In this work, we focus on accuracy.

We solve this problem using Gaussian process regression (GPR), a model-free machine learning method.
Our idea is shown in Fig.~1(d). 
The ODMR spectrum and the true magnetic field in Fig.~1(c) are used for training as the input variable $\bm{x}_i$ and output variable $y_i$, respectively, and the function $f(\bm{x})$ is obtained to estimate the magnetic field from an arbitrary spectrum $\bm{x}$.
GPR is a flexible and nonparametric machine-learning protocols commonly applied to function estimation~\cite{mackay1998,Rasmussen2004,Yukito2010}.
It is characterized by the kernel function $k(\bm{x},\bm{x}')$, which represents the degree of agreement between two input variables $\bm{x},\bm{x}'$.
We adopt the most widely used kernel function
\begin{eqnarray}
k(\bm{x},\bm{x}') = \exp(-\theta||\bm{x}-\bm{x}'||^2),
\end{eqnarray}
which is known as the squared exponential kernel.
The only two hyperparameters are the variable $\theta$ and the noise of the acquired data $\beta^{-1}$ (Method).
For a robust analysis, we normalize the contrast of the ODMR spectrum and use the data after taking the first derivative for frequency as the input variable.

The red curve in Fig.~2(a) shows the prediction of the magnetic field using the training data itself as the input variable $\bm{x}'$ for checking the regression process. The vertical axis is the difference between the repredicted magnetic field and the true magnetic field. 
The light purple area is the standard deviation of the reprediction, which indicates the uncertainty of the GPR prediction of the magnetic field from the ODMR spectrum. 
The repredicted field agrees with the true field within a standard deviation range of approximately $\pm$10~$\mu$T.

To confirm that GPR can work for quantum sensing, we deduce the estimation accuracy for a test data set not used for training. The procedure corresponds to the evaluation of generalization performance in machine learning. 
Figure~2(b) shows the magnetic field predicted by the GPR from the test data reacquired by the NDE on the cover glass as the red circle, indicating that 
the predicted field is perfectly consistent with the true one regarding the z-direction.
Figure 2(c) shows the difference between predicted and true values as red circles more quantitatively.
The predicted values encompass the true values within their statistical uncertainty for a wide magnetic field range.
Taking the case mentioned above as an example, when the true value is  547.1$\pm$0.12~$\mu$T, machine learning yields 556.1$\pm$9.5~$\mu$T.
On the other hand, the fields estimated from the model fitting (blue cross) deviate from the true ones statistically and systematically. 
These results demonstrate our approach's robust and superior accuracy compared to the model-based method.

We then verify whether GPR works for NDE on different materials. 
As the training data, we use the data of the NDE on the cover glass as before. 
Figure~2(d) shows the corresponding result obtained from the NDE on the silicon wafer.
Again, the GPR yields a statistically more reliable estimation than the model.
We note that the magnetic field derivation is more subtle in this silicon case than in the above glass case because the frequency dependence of the microwave antenna, which is difficult to be determined experimentally, is more delicate (Supplementary Information).
This confirms that GPR is applicable even when it is impossible to prepare training data, such as NDE spread over magnetic materials.

So far, we have discussed only the case of applying a magnetic field in the z-direction.
Despite the isotropic distribution of NDs over the surface in the xy plane, the ODMR spectrum can be anisotropic concerning the field direction for several reasons, such as the numerical aperture of the objective lens (Supplementary Information).
Figure~2(b) also depicts the predicted magnetic field when applying the magnetic field in the x-, y-, and z-directions.
The data with the magnetic field applied in the z-direction is used as the training data for all predictions.
The results systematically deviate from the true magnetic field when applying the field in the x- and y-directions, different from the training data.
The observed anisotropy is profitable because it suggests vector field sensing with NDE.
In principle, the contribution of each directional component is tunable by changing the objective lens' numerical aperture or the direction of the excitation light's linear polarization; The magnetic field vector can then be determined by combining these data. 
Importantly, our finding directly points to improving the existing model.

We can easily suppress this direction dependence to a negligible amount by applying a bias magnetic field in the z-direction.
We measure the magnetic field generated by the current flowing through the copper wire under the cover glass as a demonstration. Figure~3(a) shows the magnetic field distribution when  800~mA is applied to the copper wire with a bias field of 912.8~$\mu$T in the z-direction. We observe that the magnetic field becomes larger as the position gets closer to the copper wire, consistent with Ampere's law. For a more quantitative evaluation, the average value of the magnetic field in the y-direction is shown in Fig.~3(b).
The fitting results based on Ampere's law plus the bias field are consistent with the experimental results within the error bars.

Finally, we evaluate the accuracy and sensitivity achieved in our method when the field is applied in the z-direction.
We calculate the difference between the true magnetic field and the predicted value from the test data for each pixel (17$\times$22~pixels, 1~pixel$~\approx 18~\mu$m$^2$), as well as the standard deviation $\sigma$. 
The integration time $t$ dependence is also obtained for each field. In this analysis, the integration time of both training and test data is varied similarly. As shown in Fig.~4(a), as the integration time for the test data increases, the standard deviation decays. The accuracy $\zeta$ and sensitivity $\eta$ are obtained by fitting $\sigma(t) = \eta t^{-0.5} + \zeta$, taking into account the shot noise due to photon counting.
Note that the accuracy $\zeta$ is less than the above uncertainty of the function estimation ($\sim 10\ \mu$T ), since $\zeta = \lim_{t\to\infty} \sigma(t)$.

The magnetic field dependence of the accuracy and the sensitivity are shown in Figs.~4(b) and 4(c), respectively. The better the accuracy and sensitivity are, the lower these values are.
The accuracy [Fig.~4(b)] degrades at low field ($<500~\mu$T) and high field ($>2000~\mu$T). 
These behaviors are inevitable due to the nature of the NV center. 
There is only a tiny change in the spectrum at low fields [see Fig.~1(c)] because splitting the ODMR spectrum due to the lattice strain obscures the magnetic field dependence~\cite{Maze2011}.
The accuracy is not good at high fields due to the decrease in the ODMR contrast because of the random distribution of NV axes in NDE.
The sensitivity [Fig.~4(c)] also shows this tendency at high fields. The sensitivity at low fields is almost zero, which is simply an artifact due to the low resolution for integrated time [see Fig. 4(a)].
As a result, we achieve the best accuracy of about 1.8~$\mu$T under the field of 1000--1500~$\mu$T with a sensitivity of $\sim 70$~$\mathrm{\mu T/\sqrt{Hz}}$.

In summary, we demonstrate the accurate local magnetometry by the machine-learning-enhanced NDE method.
The obtained high accuracy and sensitivity are sufficient for the magnetic imaging of a few layers of the van der Waals materials~\cite{Novoselov2005,Zhang2005,Cao2018,Burch2018,Mak2019} and topological materials~\cite{Hsieh2008,Hasan2021}.
This method applies not only to such topics in condensed matter physics but also to arbitrary-shaped objects such as living organisms~\cite{Schirhagl2014,Sage2013}.
The present idea is further promising in applying temperature, electric field, and pressure measurements by using NDE~\cite{Acosta2010,Dolde2011,Doherty2014,Maze2011}.

\section*{acknowledgement}
We thank MEXT-Nanotechnology Platform Program ``Microstructure Analysis Platform'' for technical support.
MT and KO acknowledge financially support from MEXT-WINGS program FoPM.
MT also acknowledgements financial support from Daikin Industries, Ltd.
This work was partially supported by the Japan Society for the Promotion of Science (Nos. JP19K23424, JP19H00656, JP19H05826, 19H05822, JP20K22325, JP21K13859).

\clearpage

\section*{Method}
\subsection*{Experimental setup}
A laser beam (515~nm, 150~mW) is output through a multimode fiber and a collimator into the free space.
It is expanded by a lens pair before entering the objective lens (Mitsutoyo M-PLAN APO 100X, numerical aperture NA=0.7, magnification 100x).
The irradiated area of the excitation beam at the focal point is approximately 400~$\mu$m in diameter.
The NV center's fluorescence is acquired by a CMOS camera (basler acA720-520um, 12-bit resolution) after passing through an objective lens, a dichroic mirror, a 514~nm notch filter, a 650~nm long-pass filter, an 800~nm short-pass filter, and an imaging lens.
The magnetic field generated in the three-axes Helmholtz coil, including the geomagnetic field, is calibrated with a tesla meter (LakeShore Cryotronics F71) to an accuracy of $\pm0.12~\mu$T.

The NDE spread on the cover glass is fixed with carbon tape on the resonator microwave antenna~\cite{Sasaki2016}. ODMR measurements are performed by inputting 25~dBm microwave power to the antenna.

The ODMR spectra are acquired with a frequency resolution of 141 points equally spaced between 2720 and 2990~MHz. The exposure time is set to 8~ms, and the frequency sweep is repeated 70 times (8~ms$\times$141$\times$70 in total).

In the measurement shown in Fig.~4, a copper wire with a diameter of 50~$\mu$m is placed along the y-axis direction. A magnetic field is generated by applying a current [Fig.~4(b) inset].

\subsection*{Sample preparation}
NDs with a standard particle size of 50~nm (NDNV50nmHi10ml, Ad\'{a}mas Nanotechnologies) containing about 30 NV centers per particle are used. 
5~$\mu$g of NDs in 5~$\mu$L of water are dispersed by ultrasonication and then dropped onto cover glass (130~$\mu$m thick, Matsunami Glass Ind.,Ltd.) and dried with a spin coater rotating at 500~rpm for 2000~seconds. The thickness of NDE is obtained to be 200~nm (about 4~layers) by measuring with a profilometer (KLA-Tencor P-7). This value corresponds to about 30,000 NDs and about 1 million NV centers per image pixel (18~$\mathrm{\mu m^2}$). The fabricated NDE looks like a white film due to light scattering under an ambient light.
The NDE on silicon is spread with the spin coater at 400~rpm, and its thickness is obtained to be 1000~nm (about 20~layers).

\subsection*{Gaussian Process Regression}

GPR is a process that estimates a function for input and output variables~\cite{Bishop2006}.
As training data, we prepare $n$ pairs of input variable vector $\bm{x}_i$ and output scalar variable $y_i$.
When the input data is $\bm{x'}$, the predicted output value $f(\bm{x'})$ is given by,
\begin{equation}
f(\bm{x}') = \bm{k}(\bm{x}')^{\mathrm{T}} (K + \beta^{-1} I)^{-1} \bm{y},
\end{equation}
where $I$ is an identity matrix, $\bm{y}$ is a column vector with $y_i$ as its $i$~th entry, $\bm{k}(\bm{x}')$ is a column vector with $k(\bm{x}_i,\bm{x}')$ as its $i$~th entry, and $K$ is an $n\times n$ matrix with $K_{i,j} = k(\bm{x}_i,\bm{x}_j)$ as its $(i,j)$~th entry, $\beta^{-1}$ is the noise intensity on the output variable $y$.
In this case, standard deviation $\bm{v}(x')$ [confidence area of Fig.~2(a)] is given by
\begin{equation}
\bm{v}(x') = k(\bm{x}',\bm{x}') + \beta^{-1} -\bm{k}^{\mathrm{T}} K^{-1} \bm{k}.
\end{equation}

In this study, $\bm{x}_i$ and $y_i$ are the ODMR spectra shown in Fig.~1(c) and the magnetic field strength obtained with a tesla meter, respectively. The optimization of the hyperparameters is performed by the minimization  of five-fold cross-validation loss using the MATLAB Statistics and Machine Learning Toolbox.
According to Eq.~(2), the magnetic field $f(\bm{x}')$ is predicted by calculating the similarity between the ODMR spectrum $\bm{x}'$ and each training data.

\subsection*{Fitting Model}

We use an existing model of the direction dependence of the ODMR spectrum of the NV center~\cite{Foy2020}.
The electron spin Hamiltonian of the NV center is given by,
$
\hat{H} = D \hat{S}_z^2 + E(\hat{S}_x^2-\hat{S}_y^2) + \gamma B\hat{S}_z,
$
where $\hat{S}_{x,y,z}$ is the x,y,z component of the spin-1 operator, $D$ is the zero-field splitting, $E$ is the lattice strain, $\gamma$ is the gyromagnetic ratio of an electron spin, $B$ is the magnetic field strength. At a magnetic field strength of about a few $\mathrm{mT}$, the two resonance frequencies of the NV center can be approximated as $f_{\pm} = D\pm\sqrt{E^2 + (\gamma B)^2}$.
The resonance shape of each NV center is approximated as the Lorentzian $L(f_{mw},f_{\pm},\delta\nu_{\pm},C_{\pm}) = C/[(f_{\pm} - f_{mw})^2 + \delta\nu_{\pm}^2]$, where $f_{mw}$ is the applied microwave frequency, $f_{\pm},\delta\nu_{\pm},C_{\pm}$ are the resonance frequencies, linewidths, and ODMR contrasts, respectively.

We apply the parameters of our experimental setup to the models of previous work~\cite{Foy2020} to obtain ODMR spectrum $S(f_{mw})$ as,
\begin{equation}\label{eq:fitting_model}
S(f_{mw}) = \cfrac{\int_0^{\pi} \kappa(\theta_{NV})P(\theta_{NV}) \int_{0}^{2\pi} [1 - L(f_{mw},f_-,\delta\nu_-,C_-) - L(f_{mw},f_+,\delta\nu_+,C_+)] d\phi_{NV}\sin\theta_{NV}d\theta_{NV}}{2\pi\int_0^{\pi} \kappa(\theta_{NV})P(\theta_{NV})\sin\theta_{NV}d\theta_{NV}}
\end{equation}
where, $P(\theta_{NV}) \propto \dfrac{\pi}{12} [32 - \cos\theta_{max}(31 + \cos(2\theta_{max})) - 6\cos(2\theta_{NV})\sin^2\theta_{max} ]$ is the collection efficiency, $\kappa(\theta_{NV}) \propto (E_x^2 + E_y^2)\pi(1 + \cos^2\theta)$ is the light absorption efficiency. Equation~(\ref{eq:fitting_model}) is used to obtain ``model'' in Figs.~2(c) and (d).

\newpage
\begin{figure}[ht]
\begin{minipage}[t]{0.61\columnwidth}
\begin{center}
\includegraphics[width=.98\linewidth]{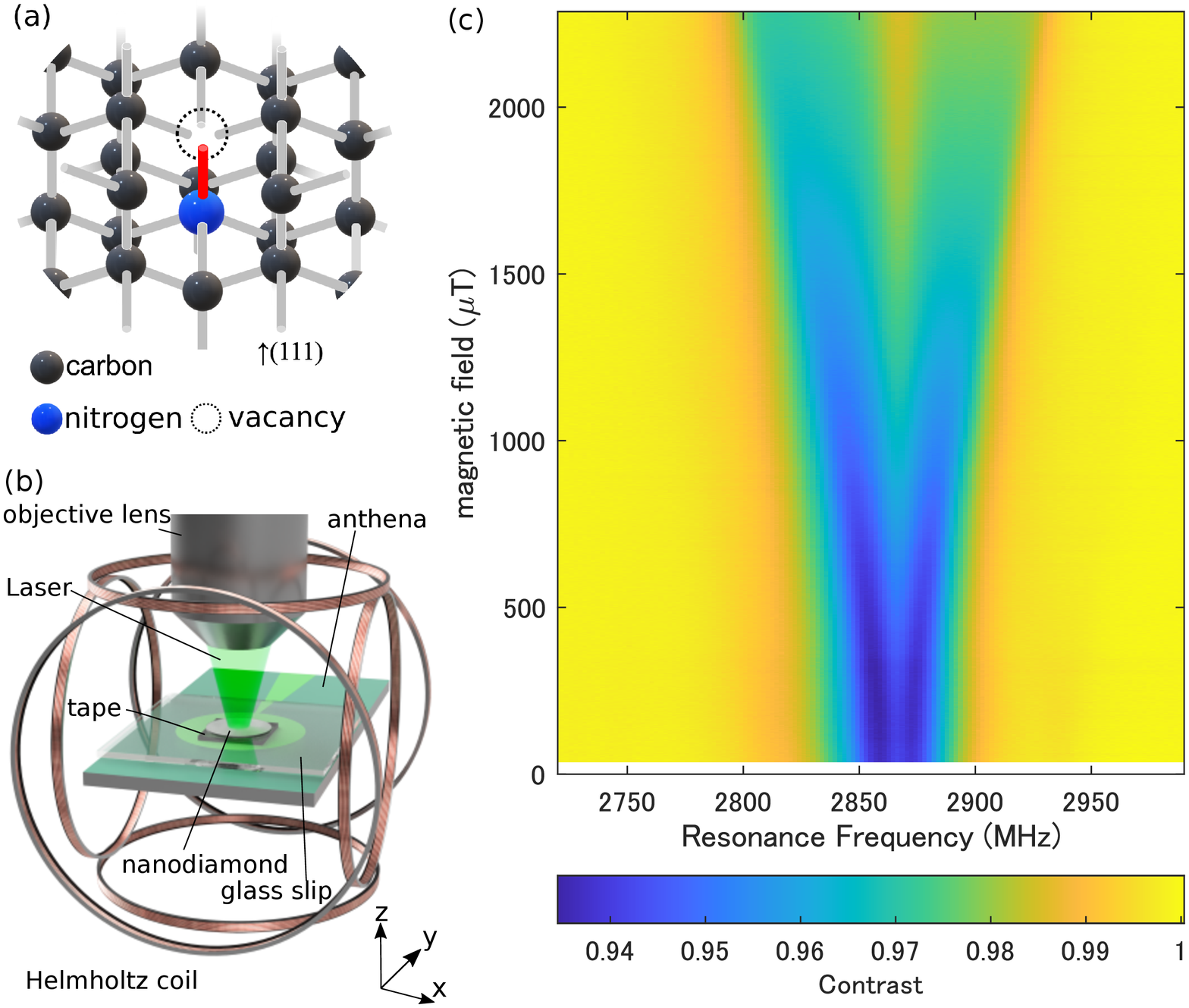}%
\end{center}
\end{minipage}%
\begin{minipage}[t]{0.38\columnwidth}
\begin{center}
\mbox{\raisebox{4mm}{\includegraphics[width=.98\linewidth]{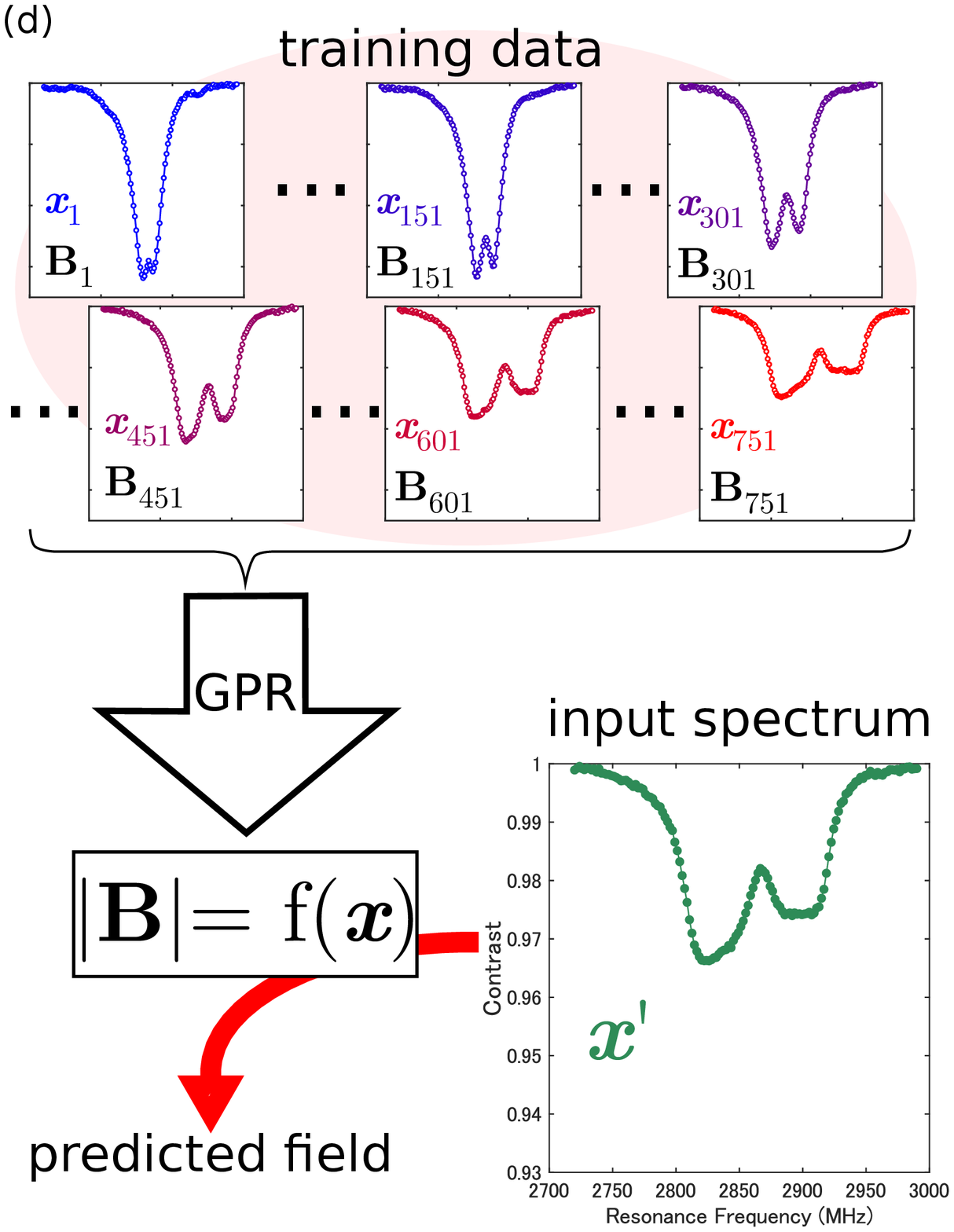}}}
\end{center}
\end{minipage}
\caption{
\textbf{Application of machine learning to diamond quantum sensors.}
(a) Schematic of a nitrogen-vacancy (NV) center in diamond.
(b) Experimental setup. The optical axis is the z-axis, and the NDE is spread on the surface in the xy-plane.
(c) Experimentally obtained ODMR spectra of NDE as functions of the microwave frequency and the magnetic field.
The true magnetic field is measured using a tesla meter.
(d) Schematic of our machine learning method.
The ODMR spectrum and the true magnetic field of (c) are used as the input vector $\bm{x}_i$ and output scalar $y_i$ for training.
Using GPR, a function is obtained from the training data to predict the magnetic field strength $|\mathrm{\bm{B}}| = f(\bm{x}')$ from an unknown spectrum $\bm{x}'$.
}
\label{f1}
\end{figure}

\newpage
\begin{figure}[th]
\includegraphics[width=.7\linewidth]{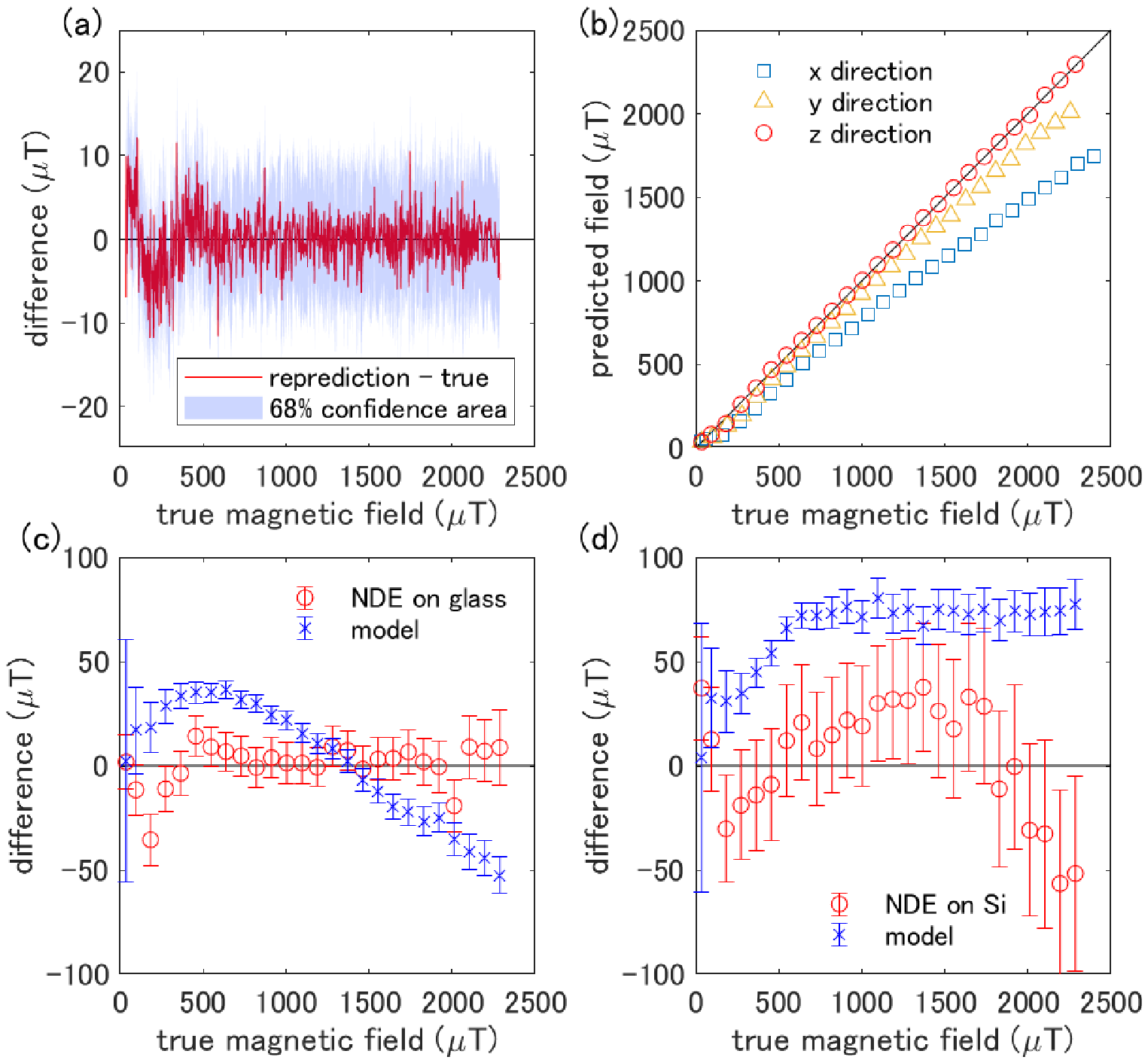}
\caption{
\textbf{Performance evaluation of GPR and comparison with the physical model.}
(a) Benchmark of reprediction by GPR. The horizontal axis is the true magnetic field, and the vertical axis is the difference between the repredicted field and the true field. The light purple region is the standard deviation interval. 
(b) Dependence of the predicted magnetic field on the magnetic field directions. The black line is the ideal value where the predicted and true magnetic fields perfectly match for the z-direction. 
(c)(d) The detail of the prediction accuracy for the NDE on the cover glass (c) and on the silicon. We use only the training data when the field is applied to the z-direction.
The difference between GPR predicted and true field shows as Red circle. The fitting result using the physical model shows as blue cross.
These errorbars depict 68\% confidential interval. 
Note that we have corrected the data for NDE on silicon regarding the heating of the material (see Fig.~S2 for detail).
}
\label{f2}
\end{figure}

\newpage
\begin{figure}[ht]
\includegraphics[width=.8\linewidth]{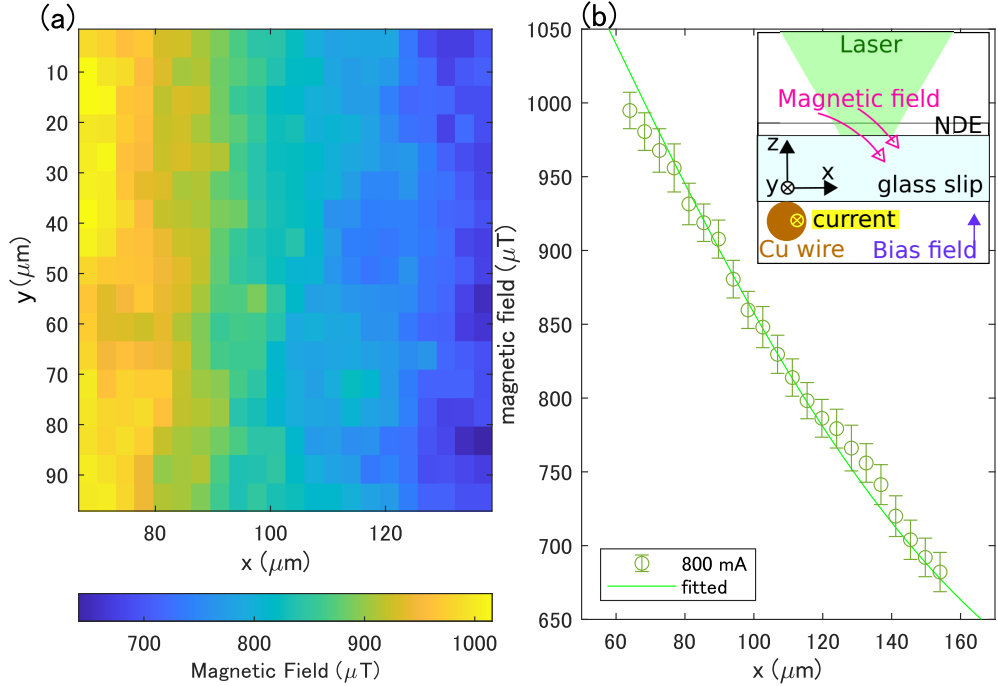}
\caption{
\textbf{Machine-learning-enhanced magnetic field imaging.}
(a) Magnetic field distribution when a current of 800~mA is applied to the copper wire, which is placed along the y-direction. The horizontal axis is the distance to the x-direction when the copper wire is placed at $x = 0\ \mu$m.
(b) The average magnetic field value against the y-axis direction in (a). The solid curve is the result of the fitting based on Ampere's law. (inset) Measurement configuration, where the bias field applied in the z-direction and the field generated by the current through the wire are simultaneously felt by the NDE.
}
\label{f3}
\end{figure}

\newpage
\begin{figure}[ht]
\includegraphics[width=.8\linewidth]{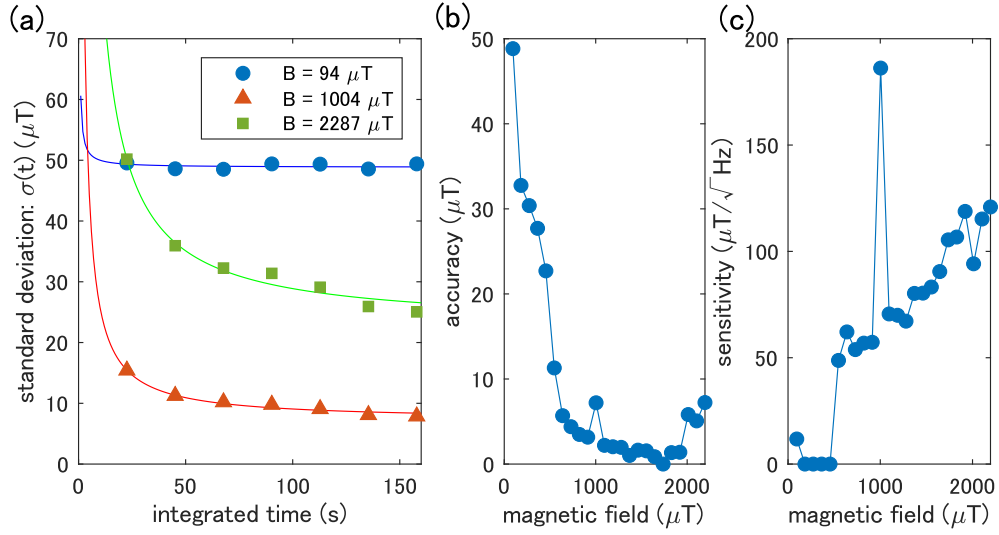}
\caption{
\textbf{Accuracy and sensitivity of machine-learning-enhanced quantum sensing.}
(a) Standard deviation of the difference between the true and the predicted magnetic field from the test data. The horizontal axis is the measurement time for test data.
(b,c) Magnetic field dependence of accuracy and sensitivity obtained by fitting of (a). 
The large outlier near 1000~$\mu$T is due to analysis error in terms of low signal-to-noise ratio (Supplementary Information).
}
\label{f4}
\end{figure}

\setcounter{figure}{0}
\setcounter{equation}{0}

\renewcommand{\thefigure}{S\arabic{figure}}
\renewcommand{\theequation}{S\arabic{equation}}
\renewcommand{\thetable}{S\arabic{table}}

\newpage

\section*{supplementary information}

\subsection*{Statistical comparison between GPR and physical model}

We discuss the statistical comparison of accuracy between our GPR (machine-learning-based) approach and the physical model~\cite{Foy2020}.

\begin{figure}[ht]
\includegraphics[width=\linewidth]{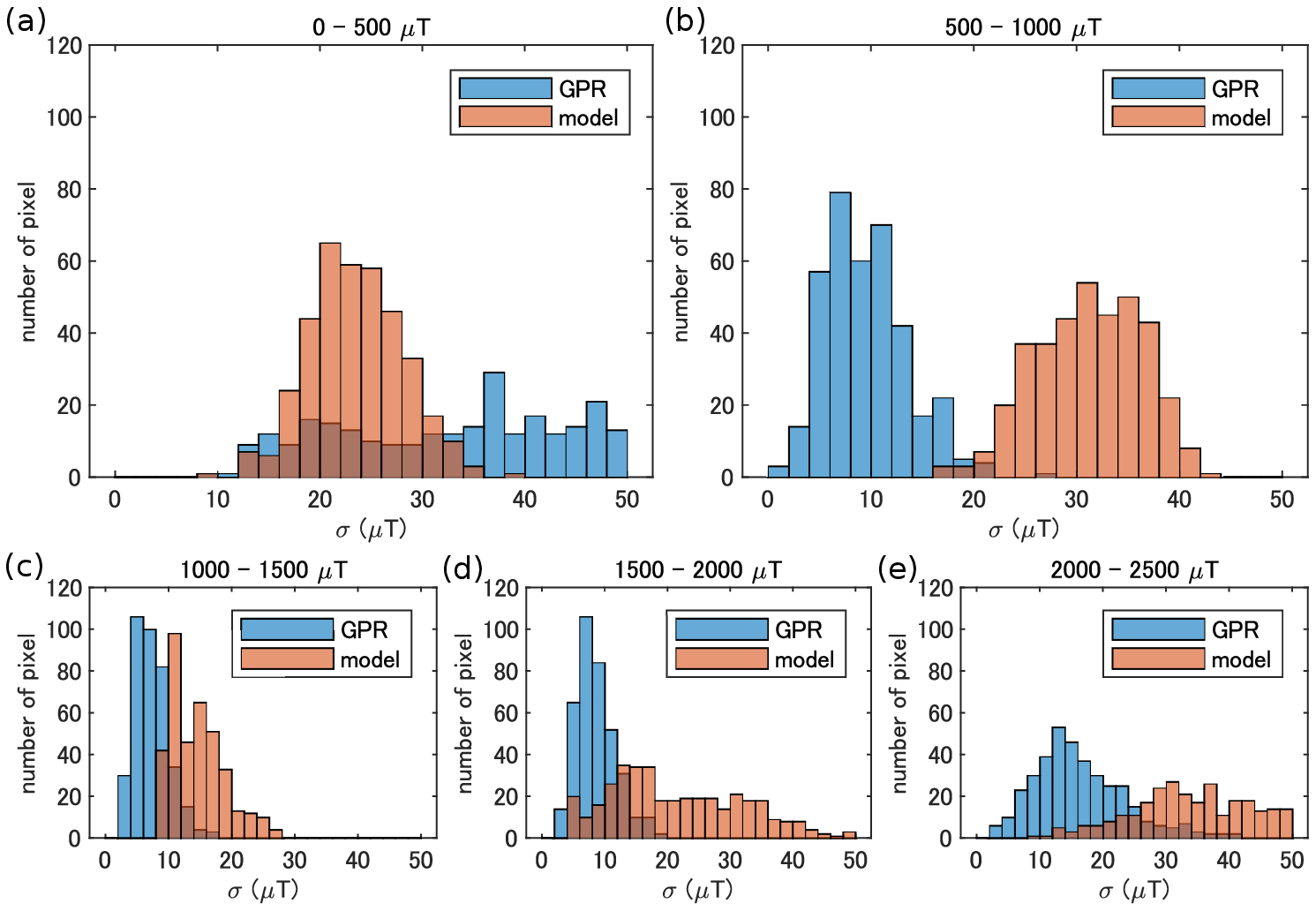}
\caption{
Standard deviation of the difference between the true and the predicted magnetic field at all pixels.
The measurement accuracy increases as the histogram shifts to the left of the graph.
The magnetic field ranges are (a) 0--500~$\mu$T, (b) 500--1000~$\mu$T, (c) 1000--1500~$\mu$T, (d) 1500--2000~$\mu$T, and (e) 2000--2500~$\mu$T.
}
\label{histogram}
\end{figure}

As shown in Fig.~2 of the main text, the difference between the true and the predicted magnetic field is analyzed for all pixels.
Figures~\ref{histogram}(a)--\ref{histogram}(e) show the histograms of standard deviations obtained for the magnetic field range with every 500~$\mu$T.
The measurement accuracy increases as the standard deviation becomes smaller and shifts to the left of the graph.
GPR has better accuracy than the physical model except for the lowest magnetic field range (0--500~$\mu$T) [Fig.~\ref{histogram}(a)].
The best accuracy is obtained at 1000--1500~$\mu$T [Fig.~\ref{histogram}(c)], consistent with Fig.~4(b) in the main text.
These results support that GPR is superior for accurate sensing in wide magnetic field ranges.

Why is the physical model more accurate only in the lowest field  0--500~$\mu$T shown in Fig.~\ref{histogram}(a)?
GPR is not suitable since there is only a slight change in this region [Fig.~1(c) in the main text].
On the other hand, the ODMR spectrum is not broad in such a low field, which reduces the uncertainties in the physical model fitting. 
However, in the higher magnetic field, the directional dependence of the NV center and the frequency dependence of the microwave antenna significantly affect the spectrum, which results in lower accuracy in the model than GPR, as we discuss later.

\subsection*{Effects of substrates on accuracy}

We discuss the causes of accuracy suppression when training data obtained from NDE on cover glass is applied to sensing with NDE on silicon.
The glass and silicon are different in thermal conductivity, electrical conductivity, and dielectric constant.
This fact raises two main effects: the heating by laser and microwave and the frequency characteristics of the microwave antenna.

First, the degree of heating by the laser and microwaves is different between the two materials, resulting in a temperature difference in the NDE on them.
It shifts the center of the ODMR spectrum with the coefficient of $-74$~kHz/K~\cite{Acosta2010}.

\begin{figure}[ht]
\begin{center}
\includegraphics[width=.7\linewidth]{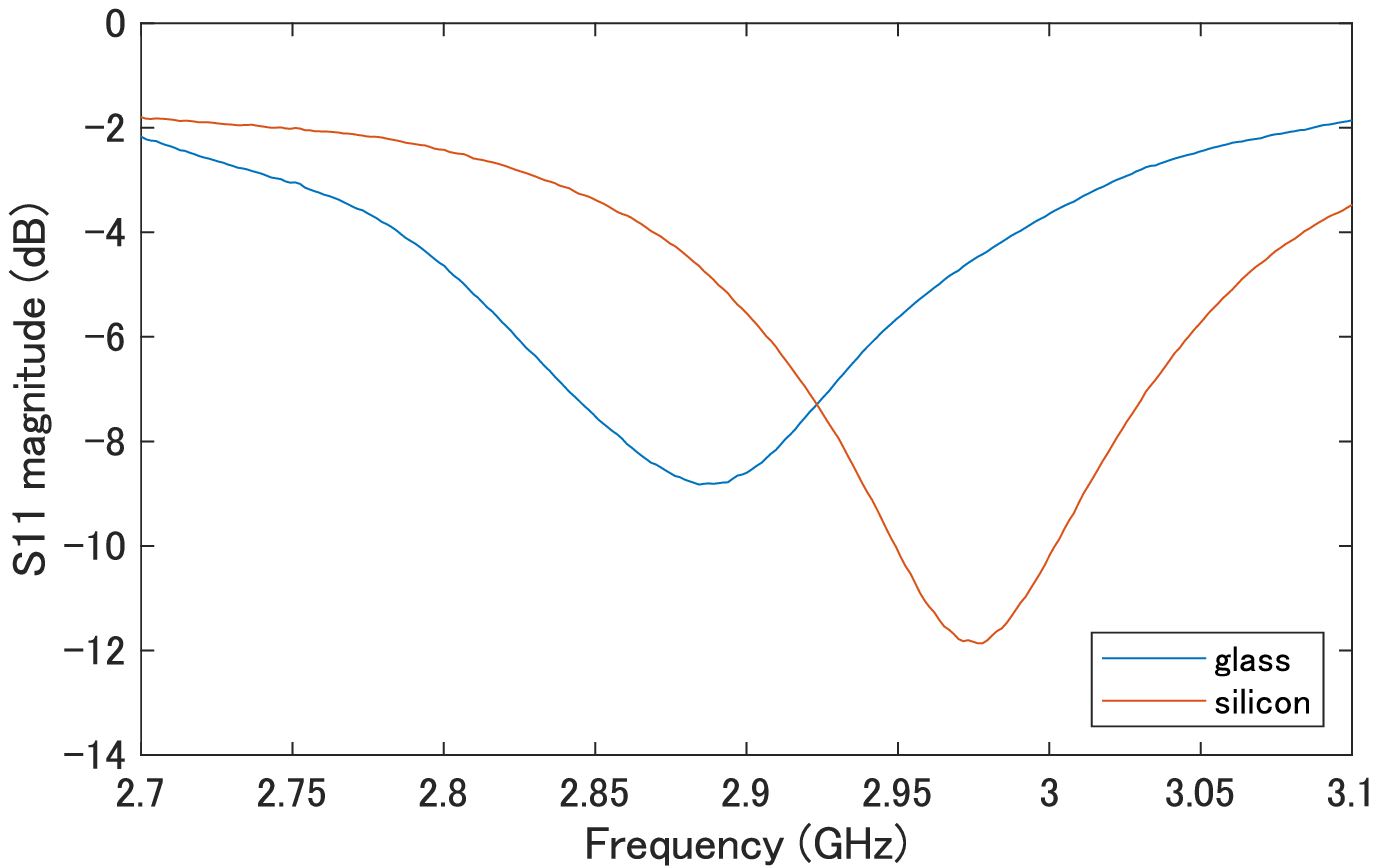}
\end{center}
\caption{
Measured S11 of our microwave antenna under each experimental condition.
A Rohde \& Schwarz FSH8 spectrum analyzer calibrated with FSH-Z28 is used.
}
\label{s11}
\end{figure}

Second, we explain the impact of the frequency characteristics of the microwave antenna. We use a broadband, large-area microwave antenna~\cite{Sasaki2016} to apply a spatially uniform microwave.
The frequency characteristics are affected by the conductivity and dielectric constant of the surrounding material.
S11 for each measurement condition is shown in Fig.~\ref{s11}.
A low S11 indicates less antenna reflection and stronger microwave radiation.
The smallest S11 frequency corresponds to the resonance frequency.
The resonance frequency on the glass is 2.88~GHz, almost the same as the center of the ODMR spectrum.
The further away from the center of the ODMR spectrum is, the weaker the microwaves become.
This leads to an underestimate of the Zeeman splitting and reduces the accuracy of the physical model.
Actually, as shown in Fig.~2(c) in the main text, the physical model gives a lower field value than the true value at high fields ($>1500~\mu$T).
On the other hand, the resonance frequency on silicon is 2.97~GHz, which is higher than that on glass.
Due to this effect, the ODMR spectrum is shifted to the  high-frequency side than it should be.
Since the shift depends on frequency, it also affects the estimation of magnetic field strength.

\begin{figure}[ht]
\includegraphics[width=\linewidth]{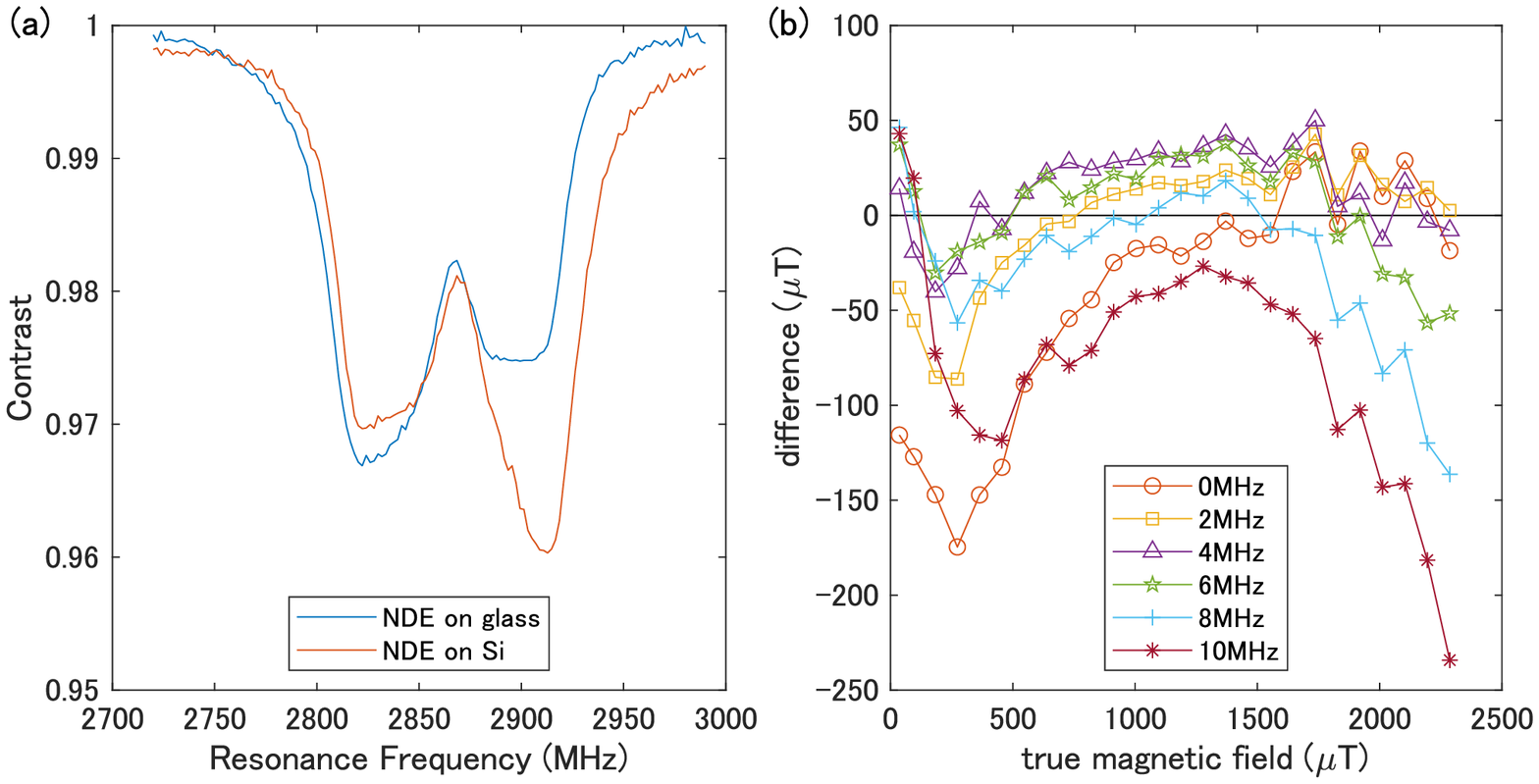}
\caption{
(a) ODMR spectra of NDE on glass and silicon. 
(b) Difference between the true magnetic field and the prediction from the test data taken on silicon with training data. The training data is taken on glass. The legends indicate the frequency shift of the training data.}
\label{S2}
\end{figure}

For example, Fig.~\ref{S2}(a) presents the ODMR spectrum obtained in a true magnetic field of 1920~$\mu$T.
The spectrum of NDE on silicon has a larger contrast on the high-frequency side than that on glass, being consistent with the characteristics of S11.
The result on glass shows that the center of the spectrum is on a lower frequency than that on silicon.
In the present experiment, black tape is placed underneath the substrate (glass or silicon) for fixing and reducing laser reflection.
The different thermal conductivity of glass and silicon contributes differently to the heating, resulting in different spectrum shifts.

We can easily correct the frequency shift caused by these sources.
Figure~\ref{S2}(b) depicts the result of the analysis by incrementally adding a shift to the frequency of the training data.
Without any correction (0~MHz), the prediction error is as large as about 150~$\mu$T. When the shift of about $+$6~MHz is incorporated, the prediction error decreases to about 50~$\mu$T.
The silicon data in Fig.~2(b) in the main text results obtained by applying this appropriate correction.

\subsection*{Physical model for the ODMR spectrum}

We describe the physical model for describing the ODMR spectrum of NDE following Ref.~\cite{Foy2020}. 
We consider the resonance frequency, the shape of the resonance, the light absorption efficiency, and the photon collection efficiency.

When the magnetic field is parallel to the NV center's symmetry axis (NV axis), the electron spin Hamiltonian of the NV center is given by,
\begin{equation}
\hat{H} = D \hat{S}_z^2 + E_{\mathrm{s}}(\hat{S}_x^2-\hat{S}_y^2) + \gamma B \hat{S}_z,
\end{equation}
where $\hat{S}_{x,y,z}$ are the x,y, and z components of the spin-1 operator, respectively, $D$ is the zero-field splitting, $E_{\mathrm{s}}$ is the lattice strain, $\gamma$ is the gyromagnetic ratio of an electron spin, and $B$ is the magnetic field strength. 
At a magnetic field strength of a few~mT, the two resonance frequencies of the NV center can be approximated as $f_{\pm} = D\pm\sqrt{E_{\mathrm{s}}^2 + (\gamma B)^2}$.

The resonance shape of a single NV center is approximated as Lorentzian $L(f_{\mathrm{mw}},f,\delta\nu,C) = C/[(f - f_{\mathrm{mw}})^2 + \delta\nu^2]$, where $f_{\mathrm{mw}}$ is the applied microwave frequency, $C$ is the contrast, and $\delta\nu$ is the linewidth.

\begin{figure*}[t]
\includegraphics[width=\linewidth]{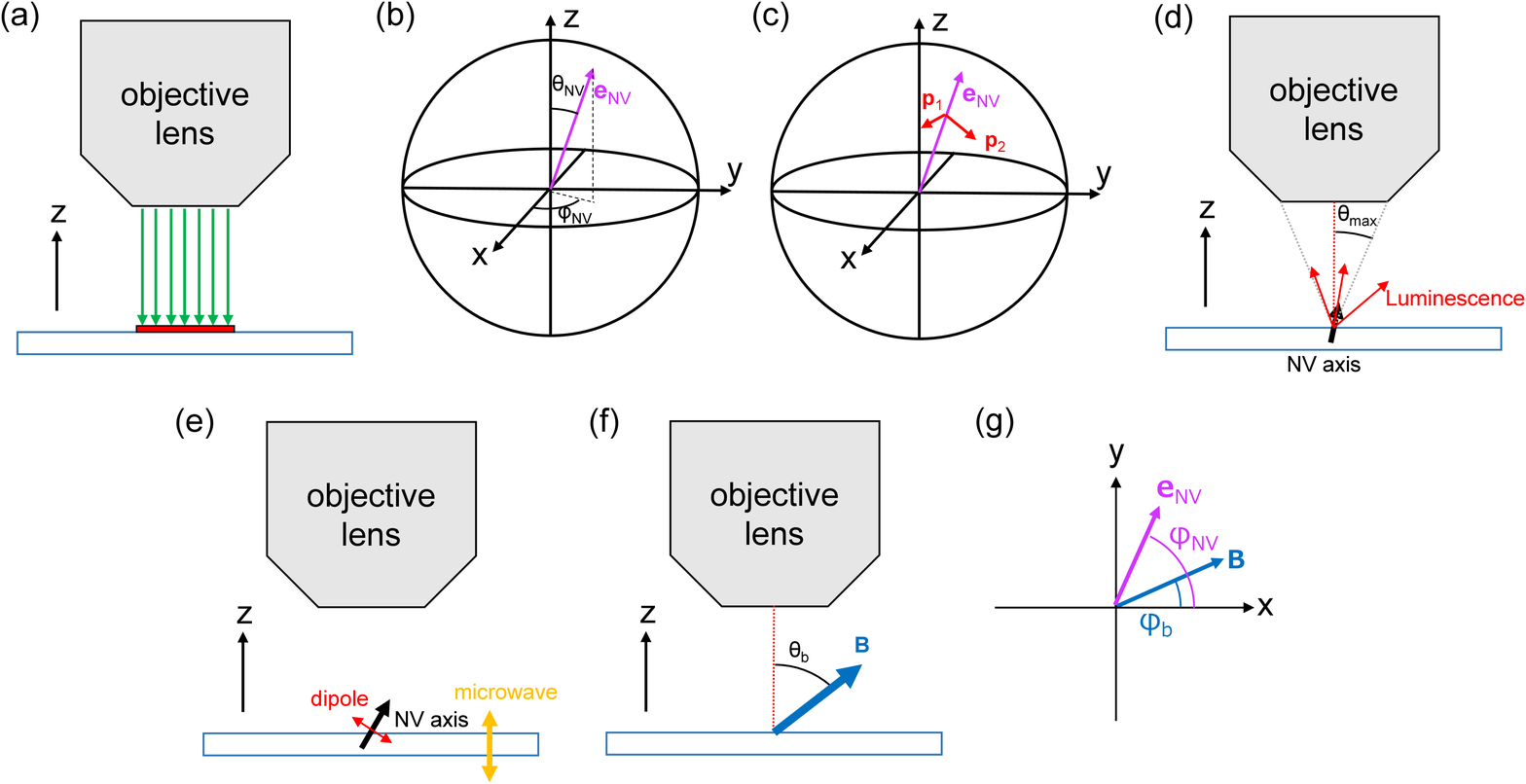}
\caption{
(a) Schematic of the K\"{o}fler type illumination. The excitation light (green arrows) is always irradiated perpendicular to the NDE film (red square) at any position.
(b) Unit vector parallel to the NV axis. $\theta_{\mathrm{NV}}$ and $\varphi_{\mathrm{NV}}$ are defined.
(c) NV center's electric dipole moments $\bm{p}_1$ and $\bm{p}_2$. 
(d) Schematic image of luminescence from NV center. 
The objective lens collects lights within the angle $\theta_{\mathrm{max}}$, which is determined by the aperture of the objective lens.
(e) Schematic of the microwave field and the NV center's spin dipole.
(f) Tilt angle of the magnetic field $\theta_b$ with respect to the optical axis.
Only the case of $\theta_b = 0$ is considered in this model.
(g) Magnetic field direction in the xy-plane when $\theta_b\neq0$.
}
\label{FigSetup}
\end{figure*}

We carefully define the experimental situation to account for absorption and collection efficiency.
Figires~\ref{FigSetup}(b) and (c) define the unit vector  parallel to the NV axis $\bm{e}_{\mathrm{NV}}$ with the polar angle $\theta_{\mathrm{NV}}$ and the azimuthal angle $\varphi_{\mathrm{NV}}$ in the Cartesian coordinate system with the optical axis in the z-axis.
For simplicity, we only show the case where the magnetic field is applied in the z-axis [Fig.~\ref{FigSetup}(a)], as in previous study~\cite{Foy2020} and training data measurements.

The light absorption efficiency is proportional to the square of the inner product of the optical transition dipole moment of the NV center and the electric field of the excitation light.
The optical transition dipole of the NV center originates from the mixed orbital~\cite{Maze2011} and has components parallel to the two orthogonal unit vectors $\bm{p}_1$ and $\bm{p}_2$.
Hereafter, we treat the case as a classical electric dipole moment for simplicity.
Despite the $C_{3 \upsilon}$ symmetry of the NV center, the NV center exhibits an axisymmetric response to linear polarization~\cite{Alegre2007}.
Assuming that dipole moments are equivalent in the any perpendicular direction to the NV axis, we define them as [Fig.~\ref{FigSetup}(c)],
\begin{align}
    \bm{p}_1 = \frac{\bm{e}_{\mathrm{NV}}\times\bm{e}_z}{|\bm{e}_{\mathrm{NV}}\times\bm{e}_z|} =& \sin\varphi_{\mathrm{NV}}\bm{e}_x - \cos\varphi_{\mathrm{NV}}\bm{e}_y,\\
    \bm{p}_2 = \frac{\bm{e}_{\mathrm{NV}}\times\bm{p}_1}{|\bm{e}_{\mathrm{NV}}\times\bm{p}_1|} =& \cos\theta_{\mathrm{NV}}(\sin\varphi_{\mathrm{NV}}\bm{e}_x + \cos\varphi_{\mathrm{NV}}\bm{e}_y) - \sin\theta_{\mathrm{NV}}\bm{e}_z,
\end{align}
where $\bm{e}_x,\bm{e}_y$, and $\bm{e}_z$ are the unit vectors parallel to x-, y-, and z-axis, respectively. 
The electric field $\bm{E}$ only in the x and y directions is applied to the NV center with the K\"{o}fler type illumination [Fig.~\ref{FigSetup}(a)].
We get the absorption efficiency $\kappa$ as,
\begin{eqnarray}
\kappa(\theta_{\mathrm{NV}}) &\propto& \int^{2\pi}_{0} (|\bm{p}_1 \cdot \bm{E}|^2 + |\bm{p}_2 \cdot \bm{E}|^2) d\varphi_{\mathrm{NV}} 
\nonumber \\
&=& (E_x^2 + E_y^2)\pi(1 + \cos^2\theta_{\mathrm{NV}}),
\end{eqnarray}
where $\bm{E}=E_x\bm{e}_x+E_y\bm{e}_y$.
The absorption efficiency is higher when the NV axis is parallel to the optical axis.
This means that photons coming from the direction of the NV axis are absorbed better.

The photon collection efficiency is also determined by the electric dipole moment~\cite{Horowitz2012}.
The energy of light emitted from a point dipole $\bm{p}$ to a position $\bm{r}$ well away from its wavelength is,
\begin{equation}
|\bm{S}| \propto | \bm{e}_r\times\bm{p} |^2/r^2,
\end{equation}
where $\bm{r}=r\bm{e}_r= r(\cos{\phi_r}\sin{\theta_r}\bm{e}_x
+ \sin{\phi_r}\sin{\theta_r}\bm{e}_y
+ \cos{\theta_r}\bm{e}_z)$.
The light is more likely to be emitted perpendicular to $\bm{p}$, i.e., in the direction of the NV axis.
The measurable angle of an objective lens is determined by NA$=\sin{\theta_{\mathrm{max}}}$ [Fig.~\ref{FigSetup}(d)].
The light collection efficiency is obtained as,
\begin{eqnarray}
P(\bm{p}) \propto \int_{0}^{\theta_{\mathrm{max}}} \int_{0}^{2\pi} 
|\bm{S}| r^2 \sin{\theta_r} d\phi_r d\theta_r.
\label{eqpp}
\end{eqnarray}
By replacing the dipole direction with the tilt of the NV center, we can integrate Eq.~(\ref{eqpp}) as,
\begin{widetext}
\begin{equation}
P(\theta_{\mathrm{NV}}) \propto \dfrac{\pi}{12} [32 - \{31 + \cos(2\theta_{\mathrm{max}})\}\cos\theta_{\mathrm{max}} - 6\cos(2\theta_{\mathrm{NV}})\sin^2\theta_{\mathrm{max}} ].
\end{equation}
\end{widetext}

Putting all the components together, we obtain the result for ODMR spectrum $S(f_{\mathrm{mw}})$ as,
\footnotesize
\begin{equation}\label{eq:spectrum}
S(f_{\mathrm{mw}}) = \cfrac{\int_0^{\pi} \kappa(\theta_{\mathrm{NV}})P(\theta_{\mathrm{NV}}) \int_{0}^{2\pi} [1 - L(f_{\mathrm{mw}},f_-,\delta\nu_-,C_-) - L(f_{\mathrm{mw}},f_+,\delta\nu_+,C_+)] d\varphi_{\mathrm{NV}}\sin\theta_{\mathrm{NV}}d\theta_{\mathrm{NV}}}{2\pi\int_0^{\pi} \kappa(\theta_{\mathrm{NV}})P(\theta_{\mathrm{NV}})\sin\theta_{\mathrm{NV}}d\theta_{\mathrm{NV}}}
\end{equation}
\normalsize
where $f_{\pm}, \delta\nu_{\pm}, C_{\pm}$ are the resonance frequencies, linewidths, and ODMR contrasts of the two resonances ($\pm$), respectively.
Equation~(\ref{eq:spectrum}) agrees with the previous model~\cite{Foy2020}.
Equation~(\ref{eq:spectrum}) is used for fitting in Figs.~2(c) and 2(d) of the main text.

Figures~S5 (a) and (b) show two examples of the fitting to the physical model.
While the estimated magnetic field deviates from the true value, the model seems to reproduce the experimental spectrum at a low field reasonably ($B = 547.1\pm0.12$~$\mu$T) [Fig.~S5(a)]. However, this is not the case at a high field ($B = 2103 \pm 0.12$~$\mu$T) [Fig.~S5(b)].
This difference indicates that the complexity of the experimental conditions is not fully reflected in the model.
Specifically, we describe two contributions as follows.
First, as mentioned in the previous section, the frequency dependence of the microwave strength radiated from a resonator-type microwave antenna exists.
Since the resonance spectrum of an NDE is continuous over a wide range, even a small frequency dependence of the antenna distorts the spectral shape.
Second, microwave absorption is dependent on the directions~[Fig.~\ref{FigSetup}(e)].
It is easy to absorb the microwave field of the component perpendicular to the NV axis.
The contrast of the NV center, which is oriented in the same direction as the microwave field, is reduced, resulting in inaccurate magnetic field estimation.

In this physical model, the magnetic field direction is assumed to be in the optical (z-axis) direction.
If the magnetic field direction deviates from the z-axis [Fig.~\ref{FigSetup}(f)], the dependence on the azimuth angle needs to be considered [Fig.~\ref{FigSetup}(g)].
The light absorption efficiency and photon collection efficiency must be integrated against the azimuth angle at the same time with the resonance shape, making the analytical calculation more difficult.
The actual effects that can occur are discussed in the next section.

\subsection*{Directional dependence of predicted magnetic field}

We observe three-dimensional field direction dependence of ODMR spectra [see Fig.~2(b) in the main text]. There are three causes of the GPR prediction difference in the z- and xy- components and an additional one cause of the difference in the x- and y-directions. We explain these four causes in order below.

The first is the absorption efficiency of the NV center. Due to K\"{o}fler type illumination, the NV symmetry axis in the z-direction ($\theta_{NV}=0$) is easily excited.

The second is the light collection efficiency~\cite{Horowitz2012}. The objective lens has a better light collection efficiency in the z-direction, and the NV has a high photon emission probability in the direction of the NV symmetry axis. 

The third is the angular dependence of the microwave absorption efficiency, as shown in Fig.~\ref{FigSetup}(e).
The microwave field perpendicular to the NV symmetry axis drives the NV center's magnetic resonance~\cite{Alegre2007}. 
In our experiment, we apply linearly polarized microwaves in the z-direction, so it is easy to obtain ODMR contrast of the NV center in the x and y directions. This is the opposite contribution to the first and second causes.

The fourth is the difference in the absorption efficiency due to the linearly polarized component of the excitation light.
The NV center is more easily excited when its symmetry axis is perpendicular to linear polarization than in the parallel case.
Since the excitation light used in this experiment is elliptically polarized, the linearly polarized component of the excitation light causes a bias in the estimated magnetic field in the x- and y-directions.

All these effects contribute to the results.
In Fig.~2(b) in the main text, GPR predicts lower values than the true magnetic field when the magnetic field is applied in the x- and y-directions.
The contribution of the third and fourth causes is estimated to be significant under our experimental conditions.
We also confirmed that the difference for the x- and y-directions is modulated by the light polarization (data not shown).

\subsection*{Additional data for magnetic field imaging}
We adopt Ampere's law for fitting the magnetic field distribution in Fig.~3(b). The only fitting parameters are the positions of the copper wire in the x- and z-directions. In addition to the data shown in Fig.~3(a) of the main text, we measure magnetic field distributions at three different current values. Figures~\ref{MagImage}(a), \ref{MagImage}(b), and \ref{MagImage}(c) show the obtained magnetic field distributions. At all current values, the magnetic fields are larger at the position closer to the copper wire. The averaged results in the y-direction are summarized in Fig.~\ref{MagImage}(e). All the experimental results are consistent with Fig.~3(b) in the main text; We can perfectly reproduce them using the exactly same wire position.

Figure~\ref{MagImage}(e) tells that the data of different current values cross each other. This is because the direction of our bias magnetic field and the magnetic field generated by the current are opposite, so they cancel each other.

If we look carefully at the magnetic field distribution, we identify that there are spatial fluctuations.
It is due to the speckle of the excitation light passing through the multimode fiber. 
Figure~\ref{MagImage}(d) shows the photoluminescence intensity distribution of the NV center. 
The area with low intensity is the area with low excitation light intensity. 
The signal-to-noise ratio in such regions is low, partially degrading the magnetic field accuracy. 
Note that we have confirmed that such inhomogeneity does not affect significantly impact the machine-learning outcome.

\newpage

\newpage

\newpage

\newpage

\newpage

\begin{figure*}[ht]
\includegraphics[width=\linewidth]{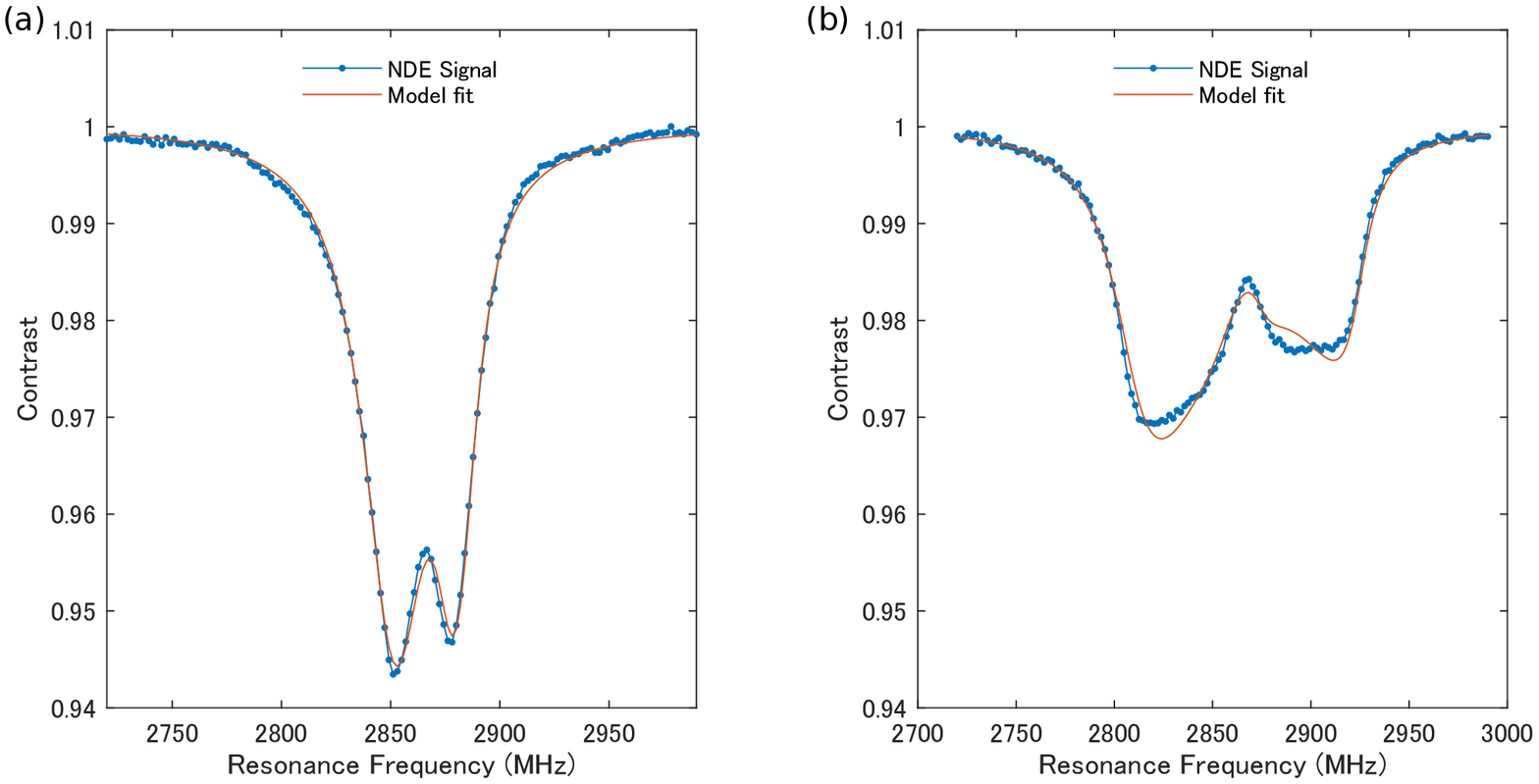}
\caption{Examples of fitting with the physical model. 
(a) The ODMR spectrum at true magnetic field $B = 547.1\pm0.12$~$\mu$T.
The estimated value is $582.2 \pm 4.6$~$\mu$T. 
(b)The ODMR spectrum at true magnetic field $B = 2103 \pm 0.12$~$\mu$T. 
The estimated value is $2062 \pm 8$~$\mu$T.}
\label{S4}
\end{figure*}

\newpage
\begin{figure*}[ht]
\includegraphics[width=\linewidth]{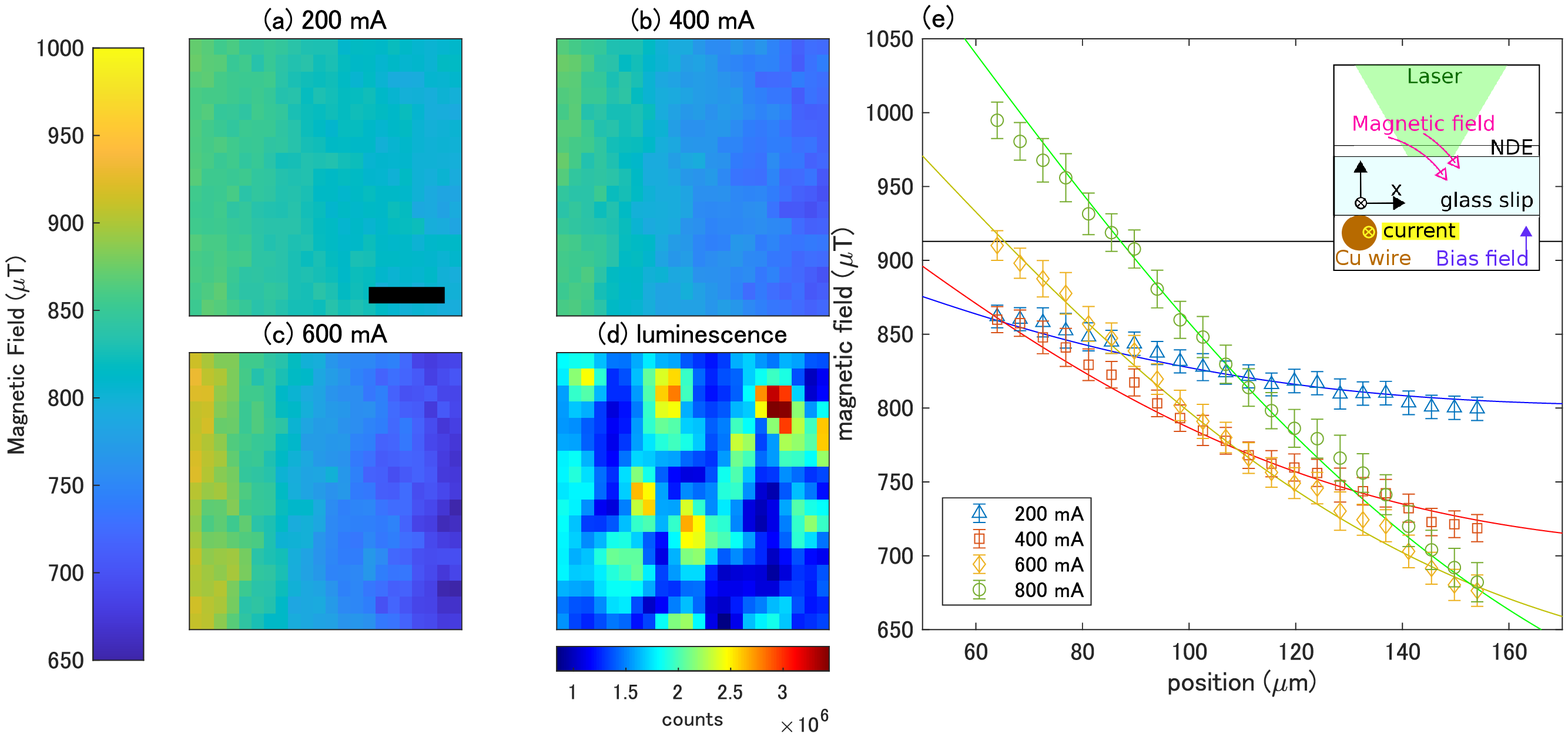}
\caption{
Magnetic field distributions at currents of (a) 200~mA, (b) 400~mA, and (c) 600~mA. Scale bar is 20~$\mu$m. 
(d) Photoluminescence intensity distribution. 
(e) Magnetic field distribution averaged in the y-direction for 200~mA, 400~mA, 600~mA, and 800~mA.
The data for 800~mA is the same as shown in Fig.~3(b) in the main text.
The colored solid lines are the fitting based on Ampere's law. The horizontal black solid line indicates the bias field strength. 
(inset) Measurement configuration, where the bias field applied in the z-direction and the field generated by the current through the wire are simultaneously felt by the NDE.}
\label{MagImage}
\end{figure*}

\clearpage
\bibliography{FNDML.bib}

\end{document}